\documentclass[%
 reprint,
nofootinbib,
 amsmath,amssymb,
 aps,
]{revtex4-1}

\usepackage{rotating}
\usepackage{hyperref}
\hypersetup{
    colorlinks=true,
    linkcolor=blue,
    filecolor=blue,      
    urlcolor=blue,
    pdftitle={Sharelatex Example},
    bookmarks=true,
    pdfpagemode=FullScreen,
}
\usepackage{tikz}
\usetikzlibrary{arrows}
\usetikzlibrary{positioning}
\usetikzlibrary{decorations.text}
\usetikzlibrary{decorations.pathmorphing}
\usepackage{feynmp}
\usepackage{multirow}
\usepackage{graphicx}% Include figure files
\usepackage{dcolumn}% Align table columns on decimal point
\usepackage{bm}% bold math
\begin{document}
\preprint{APS/123-QED}

\title{Three-Body Decay of $\Lambda_c^{*}(2595)$ and $\Lambda_c^{*}(2625)$ with the Inclusion of Direct Two-Pion Coupling}% Force line breaks with \\

%\thanks{A footnote to the article title}

\author{A. J. Arifi$^1$}
\author{H. Nagahiro$^{1,2}$}
\author{A. Hosaka$^{1,3}$}

\affiliation{%
$^1$Research Center for Nuclear Physics (RCNP), Osaka University, Ibaraki, Osaka 567-0047, Japan \\
$^2$Department of Physics, Nara Women's University, Nara 630-8506, Japan\\
$^3$Advanced Science Research Center, Japan Atomic Energy Agency, Tokai, Ibaraki 319-1195, Japan
}%
\date{\today}% It is always \today, today,

\begin{abstract}
We investigate the three-body decays of $\Lambda_c^*(2595)$ and $\Lambda_c^*(2625)$ into $\Lambda_c \pi \pi$ by including 
a direct process going through the two-pion coupling of $\Lambda_c^*\Lambda_c \pi \pi$ in addition to the sequential processes going through $\Sigma_c(2455)$ and $\Sigma_c^*(2520)$.  
The strength of the direct coupling is related to that of the Yukawa coupling by the chiral partner structure between $\Lambda_c$'s and $\Sigma_c$'s, while the strength of the Yukawa coupling is estimated by the quark model.
It is found that the direct process gives a considerable contribution especially in the decay of $\Lambda_c^*(2625)$. 
A clear indication of the direct process is seen in an asymmetric pattern in the angular correlation between the two pions.
We discuss the usefulness of the detailed analysis of the decays for the study of the structure of $\Lambda_c^*$. 
\end{abstract}

\pacs{Valid PACS appear here}% PACS, the Physics and Astronomy
\maketitle

\section{Introduction}
Three-body decays of heavy baryons have been observed experimentally and studied theoretically by using various hadron models~\cite{Cho:1994vg,Cheng:2015naa,Arifi:2017sac,Kawakami:2018olq,Mu:2014iaa}.  
Their properties such as masses and decay widths have been extracted, and also spin and parity $J^P$ have been identified~\cite{Abe:2006rz,Artuso:2000xy,Joo:2014fka,Aaij:2017vbw}. 
For instance, the spin $J = 5/2$ of $\Lambda_c^*(2880)$ was clearly identified by the decay angular correlations~\cite{Abe:2006rz}.  
Detailed analysis of line shapes of invariant mass distributions is also useful to know the internal structures of the resonances in comparison with theoretical studies. 

\begin{figure}[b]
\begin{tikzpicture}

\draw[line width=0.7mm] (0.5,0.4) -- (3.5,0.4);
\draw[line width=0.5mm,dashed,red,->] (2.8,3.6) -- (2.8,0.6);
\draw[line width=0.5mm,dashed,red,->] (3,4.2) -- (3,0.6);
\draw[line width=0.5mm,dashed,blue,->] (3.2,3.6) -- (5,2.7);
\draw[line width=0.5mm,dashed,blue,->] (5,2.5) -- (3.2,0.6);
\draw[line width=0.5mm,dashed,blue,->] (3.2,3.6) -- (5,1.7);
\draw[line width=0.5mm,dashed,blue,->] (5,1.5) -- (3.2,0.6);
\draw[line width=0.5mm,dashed,blue,->] (3.2,4.2) -- (5,2.7);
\draw[line width=0.5mm,dashed,blue,->] (3.2,4.2) -- (5,1.7);
\node at (1.7,0.7) {$\Lambda_c (g.s.),1/2^+$};
\node at (1.7,4) {$\Lambda_c^* (2595),1/2^-$};
\node at (1.7,4.6) {$\Lambda_c^* (2625),3/2^-$};
\draw[line width=0.7mm] (0.5,3.7) -- (3.5,3.7)  ;
\draw[line width=0.7mm] (0.5,4.3) -- (3.5,4.3);
\node at (6.3,1.9) {$\Sigma_c (2455),1/2^+$};
\node at (6.3,2.9) {$\Sigma_c^* (2520),3/2^+$};
\draw[line width=0.7mm] (4.5,1.6) -- (7.5,1.6);
\draw[line width=0.7mm] (4.5,2.6) -- (7.5,2.6);

\end{tikzpicture}
\caption{\label{chiralpartner} Direct decay processes (red dashed lines) and sequential decay processes (blue dashed lines) of $\Lambda_c^*(2595)$ and $\Lambda_c^*(2625)$.}
\end{figure}
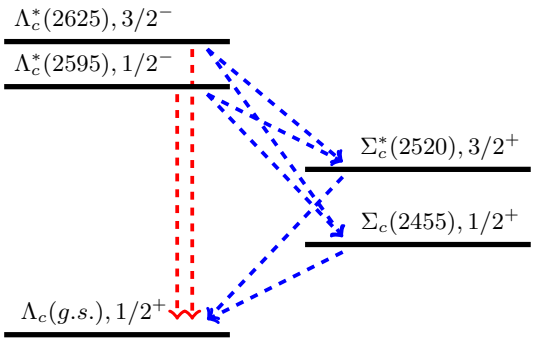

Among various charmed baryon resonances, the decay of the lowest one $\Lambda_c^*(2595)$ of $J^P = 1/2^-$ is dominated by the sequential process through $\Sigma_c(2455)$ and is well explained by identifying that resonance with a $p$-wave $\lambda$ mode excitation~\cite{Arifi:2017sac}.
Possible sequential processes are shown by blue-dashed arrows in Fig.~\ref{chiralpartner}.
The property of the next one $\Lambda_c^*(2625)$ is more interesting because the decay is dominated by a non-resonant contribution.
It was shown in our previous work that a significant amount of the non-resonant contributions was brought by the tail of the higher resonance $\Sigma_c^*(2520)$, which is not kinematically allowed as a real sequential process ~\cite{Arifi:2017sac}.  
Yet, if we take the upper bound of the current experimental data for $\Lambda_c^*(2625)$ decay width, we would expect some more contributions from other decay processes.  

One of the possible processes for the non-resonant contributions is the direct process as indicated by the red-dashed arrows in Fig.~\ref{chiralpartner}.
In this regard, recently Kawakami and Harada proposed a chiral and heavy quark classification for the four baryons, $\Sigma_c(2455, 2520)$ and $\Lambda_c^*(2595, 2625)$ \cite{Kawakami:2018olq}. 
The underlying mechanism is to identify light-diquarks in these baryons with chiral partners each other in the presence of one heavy quark.
One of the non-trivial consequences of their model is that the coupling strength of the direct process in $\Lambda_c^*(2625) \to \pi \pi \Lambda_c(g.s.)$ is related to the Yukawa coupling of $\Lambda_c^*(2625) \to \pi \Sigma_c^*(2520)$.
As a consequence, they have found that the direct process contributes a lot to the total decay width.

In this paper, we study the effect of the direct coupling of the two pions in the decays of $\Lambda_c^*(2595)$ and $\Lambda_c^*(2625)$. 
We are not only analyzing the effect in various Dalitz plots and invariant mass distributions, but also the angular correlations between the two pions. 
The presence of the direct process is clearly indicated by the asymmetric pattern in angular correlations which can be measured experimentally.

This paper is organized as follows. 
In Sec. II, we discuss the formalism.
In Sec. III, we explain the three-body decay kinematics.
In Sec. IV, we discuss the results of three-body decay with the direct process.
Finally, a summary is given in Sec. V. 
The detailed calculations of the decay amplitudes are shown in Appendix A and B.
For completeness, the results from the other quark model assignments to $\Lambda_c^*(2625)$, $\rho$ mode $(j=1)$ and $(j=2)$, are provided in Appendix C and D.

\section{Formalism}

\subsection{Chiral Partner Structures}

Chiral symmetry relates the positive and negative parity particles as chiral partners each other.
They are degenerate in the chiral limit.  
Due to the spontaneous breaking of chiral symmetry, the degeneracy is resolved, and the mass splitting occurs between the chiral partners. 

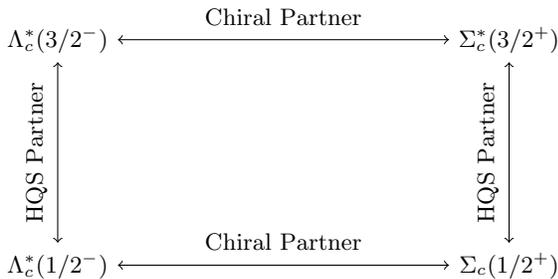
\begin{figure}[b]
\begin{tikzpicture}
\node at (1,2) {$\Lambda_c^* (1/2^-)$};
\draw[decoration={text along path, text={HQS Partner},text align={center}},decorate] (0.8,2.3) -- (0.8,4.7);
\draw[<->] (1,2.3) -- (1,4.7);
\draw[decoration={text along path, text={HQS Partner},text align={center}},decorate] (6.8,2.3) -- (6.8,4.7);
\draw[<->] (7,2.3) -- (7,4.7);
\draw[decoration={text along path, text={Chiral Partner},text align={center}},decorate] (1.8,5.2) -- (6.2,5.2);
\draw[<->] (1.8,5) -- (6.2,5);
\draw[decoration={text along path, text={Chiral Partner},text align={center}},decorate]  (1.8,2.2) -- (6.2,2.2);
\draw[<->] (1.8,2) -- (6.2,2);
\node at (1,5) {$\Lambda_c^* (3/2^-)$};
\node at (7,2) {$\Sigma_c (1/2^+)$};
\node at (7,5) {$\Sigma_c^* (3/2^+)$};
\end{tikzpicture}

\caption{\label{chiral} Chiral and heavy quark symmetry (HQS) partner classification between $\Lambda_c$'s and $\Sigma_c$'s.}
\end{figure}

The study of the chiral partner structure has been done in many cases~\cite{Kawakami:2018olq,Harada:2012dm,Ma:2015lba,Nowak:2004jg,Nowak:1992um,Suenaga:2014sga}.
Recently, Kawakami and Harada extend this idea to the light-diquarks which can be an important component of charmed baryons~\cite{Kawakami:2018olq}.
By combining them with a charm quark, they classify the charmed baryons as the chiral partners each other as depicted in Fig.~\ref{chiral} where heavy quark symmetry partners are also indicated.
We will use the same assumption in this work.
As mentioned before, one of the important implications of the chiral partner assumption is that the coupling strength of the direct process is related to the Yukawa coupling in sequential decays. 
We will discuss it in detail in the subsequent subsection. 

\subsection{Amplitudes}

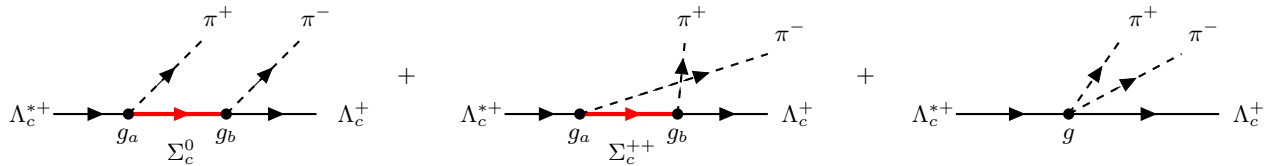
\begin{figure*}
\begin{tikzpicture}
\newcommand{\midarrow}{\tikz \draw[-triangle 45] (0,0) -- +(0.35,0);}
\newcommand{\midarrowa}{\tikz \draw[-triangle 45] (0,0) -- +(1,0);}
\newcommand{\midarrowb}{\tikz \draw[-triangle 45] (0,0) -- +(0.5,0);}

\begin{scope}[very thick, every node/.style={sloped,allow upside down}]

\node at (1,2) {$\Lambda_c^{*+}$};
\draw[thick] (1.3,2) -- node{\midarrow} (2.3,2);

\draw[ultra thick,red] (2.3,2) -- node{\midarrow} (3.6,2);

\node at (2.3,2) [circle,fill,inner sep=1.5pt]{};
\node at (2.3,1.7) {$g_a$};
\node at (3,1.5) {$\Sigma_c^{0}$};

\draw[thick] (3.6,2) -- node{\midarrow} (4.8,2);
\node at (3.6,2) [circle,fill,inner sep=1.5pt]{};
\node at (3.6,1.7) {$g_b$};

\node at (5.3,2) {$\Lambda_c^{+}$};

\draw[dashed,thick] (2.3,2) -- node{\midarrow} (3.3,3);
\node at (3.5,3.3) {$\pi^{+}$};
\draw[dashed,thick] (3.6,2) -- node{\midarrow} (4.6,3);
\node at (4.8,3.3) {$\pi^{-}$};

\node at (6,2.5) {$+$};

\node at (7,2) {$\Lambda_c^{*+}$};
\draw[thick] (7.3,2) -- node{\midarrow} (8.3,2);
\draw[ultra thick,red] (8.3,2) -- node{\midarrow} (9.6,2);
\node at (8.3,2) [circle,fill,inner sep=1.5pt]{};
\node at (8.3,1.7) {$g_a$};
\node at (9,1.5) {$\Sigma_c^{++}$};
\draw[thick] (9.6,2) -- node{\midarrow} (10.8,2);
\node at (9.6,2) [circle,fill,inner sep=1.5pt]{};
\node at (9.6,1.7) {$g_b$};

\node at (11.2,2) {$\Lambda_c^{+}$};

\draw[dashed,thick] (8.3,2) -- node{\midarrowa} (10.8,2.8);
\node at (9.8,3.3) {$\pi^{+}$};
\draw[dashed,thick] (9.6,2) -- node{\midarrowb} (9.7,3);
\node at (11.1,3.1) {$\pi^{-}$};

\node at (12.1,2.5) {$+$};

\node at (13,2) {$\Lambda_c^{*+}$};
\draw[thick] (13.3,2) -- node{\midarrow} (14.8,2);
\node at (14.8,2) [circle,fill,inner sep=1.5pt]{};
\draw[thick] (14.8,2) -- node{\midarrow} (16.8,2);
\node at (17.2,2) {$\Lambda_c^{+}$};

\draw[dashed,thick] (14.8,2) -- node{\midarrow} (16.3,2.8);
\node at (15.8,3.3) {$\pi^{+}$};
\draw[dashed,thick] (14.8,2) -- node{\midarrow} (15.5,3);
\node at (16.6,3.1) {$\pi^{-}$};
\node at (14.8,1.7) {$g$};
\end{scope}
\end{tikzpicture}
\caption{\label{3body} Feynman diagrams of three-body decay of $\Lambda_c^{*}(2595)$ into $\Lambda_c^+\pi^+\pi^-$. 
The first two diagrams represent sequential processes going through $\Sigma_c(2455)^{0}$ and $\Sigma_c(2455)^{++}$. 
The last diagram corresponds to the direct process. 
For $\Lambda_c^*(2625)$ decay, there are additional sequential processes going through $\Sigma_c^*(2520)$.}
\end{figure*}

The amplitudes for the three-body decay of $\Lambda_c^* \rightarrow \Lambda_c \pi\pi$  are described by sequential processes and direct process as shown in Fig.~\ref{3body}.
Here the direct process is the new ingredient, while the sequential process has been studied in our previous work~\cite{Arifi:2017sac}.
In Fig.~\ref{3body}, the first two diagrams describe the sequential processes and the last diagram corresponds to the direct process.

For $\Lambda_c^*(2595)$ decay, the amplitudes of the sequential processes going through $\Sigma_c(2455)$ are expressed by
\begin{eqnarray}
\mathcal{T} (\Sigma_c^{0})     &=& \frac{ G_{ab}\ \chi^\dagger_{\Lambda_c} (\boldsymbol{\sigma} \cdot {\bf p}_2)\ \chi_{\Lambda_c^*}}{m_{23} - m_{\Sigma_c^0} + i \Gamma_{\Sigma_c^0}/2},\\
\mathcal{T} (\Sigma_c^{++})   &=&  \frac{G_{ab}\ \chi^\dagger_{\Lambda_c} (\boldsymbol{\sigma} \cdot {\bf p}_1)\ \chi_{\Lambda_c^*}}{m_{13} - m_{\Sigma_c^{++}} + i \Gamma_{\Sigma_c^{++}}/2},
\end{eqnarray}
where $\chi^\dagger_{\Lambda_c}$ and $\chi_{\Lambda_c^*}$ are the spin states of $\Lambda_c(g.s.)$ and $\Lambda_c^*$ respectively, and we define $p_1, p_2$ and $p_3$ as the four-momenta of the outgoing $\pi^+, \pi^-$ and $\Lambda_c^+$ respectively as depicted in Fig.~\ref{3body}.
The invariant masses $m_{23}$ and $m_{13}$ are for the particles 2, 3 and for 1, 3 respectively.   
The coefficient $G_{ab}$ is given by
\begin{eqnarray}
G_{ab} &=& g_a g_b \sqrt{2m_{\Lambda_c^*}}\sqrt{2m_{\Lambda_c}} \label{coupling1}
\end{eqnarray}
where $g_a$ and $g_b$ are the Yukawa coupling of the first and second vertices in the sequential processes as described in Fig.~\ref{3body}.
We omit the contribution of $\Sigma_c^*(2520)$ since it is very small for the case of $\Lambda_c^*(2595)$ decay~\cite{Arifi:2017sac}.
The charmed baryon masses and decay widths are adopted from Particle Data Group (PDG)~\cite{Patrignani:2016xqp}.

For the case of $\Lambda_c^*(2625)$ decay, the amplitudes of the sequential processes going through $\Sigma_c(2455)$ and $\Sigma_c^*(2520)$ are expressed by
\begin{eqnarray}
\mathcal{T} (\Sigma_c^{0})		&=& \frac{F_{ab}\ \chi^\dagger_{\Lambda_c} (\boldsymbol{\sigma}  \cdot {\bf p}_2)}{m_{23} - m_{\Sigma_c^{0}} + i \Gamma_{\Sigma_c^{0}}/2} \nonumber\\
							&& \times \left(\boldsymbol{\sigma} \cdot {\bf p}_1\ {\bf S} \cdot {\bf p}_1  -\frac{1}{3}   \boldsymbol{\sigma} \cdot {\bf S} |{\bf p}_1|^2\right) \chi_{\Lambda_c^*}, \quad\quad\nonumber\\
\mathcal{T} (\Sigma_c^{++})   	&=&  \frac{F_{ab}\ \chi^\dagger_{\Lambda_c} (\boldsymbol{\sigma}  \cdot {\bf p}_1)}{m_{13} - m_{\Sigma_c^{++}} + i \Gamma_{\Sigma_c^{++}}/2} \nonumber\\
							&& \times \left(\boldsymbol{\sigma} \cdot {\bf p}_2\ {\bf S} \cdot {\bf p}_2  -\frac{1}{3}   \boldsymbol{\sigma} \cdot {\bf S} |{\bf p}_2|^2\right) \chi_{\Lambda_c^*},\nonumber\\
\mathcal{T} (\Sigma_c^{*0})    	&=&  \frac{F_{cd}\ \chi^\dagger_{\Lambda_c} ({\bf S} \cdot {\bf p}_2)\ \chi_{\Lambda_c^*}}{m_{23} - m_{\Sigma_c^{*0}} + i \Gamma_{\Sigma_c^{*0}}/2},\nonumber\\
\mathcal{T} (\Sigma_c^{*++})  	&=&  \frac{F_{cd}\ \chi^\dagger_{\Lambda_c} ({\bf S} \cdot {\bf p}_1) \chi_{\Lambda_c^*}}{m_{13} - m_{\Sigma_c^{*++}} + i \Gamma_{\Sigma_c^{*++}}/2},
\end{eqnarray}
where we have defined 
\begin{eqnarray}
F_{ab} &=& f_a f_b \sqrt{2m_{\Lambda_c^*}}\sqrt{2m_{\Lambda_c}},\\
F_{cd} &=& f_c f_d \sqrt{2m_{\Lambda_c^*}}\sqrt{2m_{\Lambda_c}}.
\end{eqnarray}
The values $f_a$ and $f_b$ are the Yukawa coupling of the first and second vertices in the sequential process through $\Sigma_c(2455)$ in the intermediate state, 
$f_c$ and $f_d$ are for those of $\Sigma_c^*(2520)$.
The detailed discussion about the amplitudes of the sequential processes can be found in our previous work~\cite{Arifi:2017sac}.

The direct process amplitudes adopted from Ref.~\cite{Kawakami:2018olq} are given by
\begin{eqnarray}
 \mathcal{T}_{\rm Direct} [\Lambda_c^* (2595)] &=& \frac{ g}{f_\pi}\ \bar{u}_{\Lambda_c}  (p_1+p_2)_\mu \left( \gamma^\mu + \frac{P^\mu}{M} \right)\gamma_5 u_{\Lambda_c^*},\nonumber\\ \\
\mathcal{T}_{\rm Direct} [\Lambda_c^* (2625)]&=& \frac{ f}{f_\pi}\ \bar{u}_{{\Lambda_c}} (p_1+p_2)_\mu u_{\Lambda_c^*}^\mu.
\end{eqnarray} 
where $g$ and $f$ is the coupling of the direct process for the decay of $\Lambda_c^*(2595)$ and $\Lambda_c^*(2625)$ respectively, and $f_\pi = 93$ MeV is the pion decay constant.   
The estimation of the coupling strength will be given later.
The mass and the four-momentum of the initial $\Lambda_c^*$ are denoted by $M$ an $P$.
The Dirac spinors ${u}_{\Lambda_c}$ and $u_{\Lambda_c^*}$ are for $\Lambda_c(g.s.)$ and $\Lambda_c^*$ respectively.

In our present work, we perform the non-relativistic reduction so that the amplitudes are written in terms of spin matrices as
\begin{eqnarray}
\mathcal{T}_{\rm Direct} [\Lambda_c^* (2595)] &=& \frac{G}{f_\pi}\ \chi^\dagger_{\Lambda_c} \left( \boldsymbol{\sigma} \cdot ({\bf p}_1+{\bf p}_2) \right)\chi_{\Lambda_c^* },\label{direct1}\quad\\
\mathcal{T}_{\rm Direct} [\Lambda_c^* (2625)] &=& \frac{F}{f_\pi}\ \chi^\dagger_{\Lambda_c} \left( {\bf S} \cdot ({\bf p}_1+{\bf p}_2) \right) \chi_{\Lambda_c^*}, \quad\quad \label{direct2}
\end{eqnarray}
where we have defined 
\begin{eqnarray}
G &=& g \sqrt{2m_{\Lambda_c^*}} \sqrt{2m_{\Lambda_c}},\\
F &=& f \sqrt{2m_{\Lambda_c^*}} \sqrt{2m_{\Lambda_c}}.
\end{eqnarray} 
It is found that the amplitudes shown in Eqs.~(\ref{direct1}) and (\ref{direct2}) are unique in the non-relativistic limit.
This can be understood by considering the conservation of spin and parity of the particles participating in the interaction.
The initial state  $\Lambda_c^{*} (1/2^-\ {\rm or}\ 3/2^-)$ decays into the ground state $\Lambda_c (1/2^+)$ emitting the two pions ($0^-$).
Consequently, the only possible way is that the pions carry angular momentum  $l=1$  ($p$-wave)
in order to conserve the total angular momentum during the decay process.

The $p$-wave nature of the direct process leads to $\cos \theta_{12}$ where the angle $\theta_{12}$ is defined as the angle between the two pions in the resonance's rest frame. 
We can obtain $\cos \theta_{12}$ dependence by taking the square of the amplitude in Eqs. (\ref{direct1}) and (\ref{direct2}) respectively.
This $\cos \theta_{12}$ produces the asymmetric pattern in the angle correlations between the two pions.

In order to estimate the direct process coupling strength, we use the assumption of chiral partner where it is equal to the Yukawa coupling of the second vertex in the sequential decay,
\begin{eqnarray}
\Lambda_c^*(2595): \hspace{1cm} g &=& g_b, \\
\Lambda_c^*(2625): \hspace{1cm} f &=& f_d, 
\end{eqnarray}
where $g_b$ is the coupling strength of the second vertex in the sequential decay going through $\Sigma_c (2455)$ in $\Lambda_c^*(2595)$ decay,
and $f_d$ is the coupling strength of $\Sigma_c^*(2520)$ sequential process in $\Lambda_c^*(2625)$ decay.
The values of $g_b$ and $f_d$ are extracted from the quark model for the sequential processes as in our previous work \cite{Arifi:2017sac}.

\section{Three-body Decay Kinematics}

The three-body decay width is calculated as
\begin{eqnarray}
{\rm d}\Gamma = \frac{(2\pi)^4}{2M} |\mathcal{T}|^2\ {\rm d}\Phi_3(P;p_1,p_2,p_3)
\end{eqnarray}
where ${\rm d}\Phi_3$ is an element of the three-body phase space given by
\begin{eqnarray}
{\rm d}\Phi_3 &=& \delta^4(P - p_1-p_2-p_3)\nonumber\\
	&&\times \frac{{\rm d}^3p_1}{(2\pi)^32E_1}\frac{{\rm d}^3p_2}{(2\pi)^32E_2}\frac{{\rm d}^3p_3}{(2\pi)^32E_3}. \quad \quad
\end{eqnarray}
In general, there are five independent kinematical variables to characterize three-body decays: two invariant masses and three angles, which determine the orientation of the decay plane.
In unpolarized case, the angular variables become irrelevant and then the decay depends only on the two invariant masses.

The standard way to describe the three-body decay is the so-called Dalitz plot which is represented by two invariant masses.
We can construct three invariant masses as $m_{23}^2, m_{13}^2,$ and $ m_{12}^2$ where particles 1, 2, and 3 are assigned to $\pi^+, \pi^-$ and $\Lambda_c^+$ respectively.
In the following sections, we will show two Dalitz plots in $(m_{23}^2, m_{13}^2)$ and  $(m_{12}^2, m_{13}^2)$ planes.  
The plots in $(m_{12}^2, m_{23}^2)$ plane shows similar distribution to the latter one because the masses of particle $\pi^+ (1)$ and $\pi^- (2)$ are approximately the same.

Here we are interested in the angular dependence in the three-body decay.
Previously, we ignore the angular dependence by taking the angle average  \cite{Arifi:2017sac}.
However, it turns out that the angular dependence is important and useful to see the effect of the direct process.

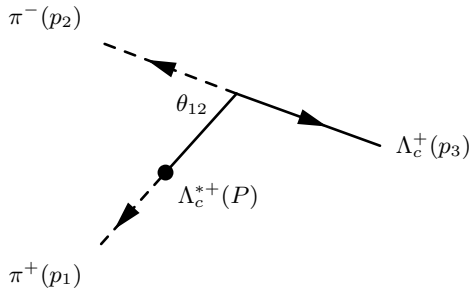
\begin{figure}[t]
 \begin{fmffile}{helicityangle}
    \begin{fmfgraph*}(120,80)
        \fmfleft{i1,i2}
        \fmfright{o1}
        \fmflabel{$\Lambda_c^{+}(p_3)$}{o1}
        \fmflabel{$\pi^- (p_2)$}{i2}     
        \fmflabel{$\pi^+(p_1)$}{i1} 
         \fmfstraight
        \fmf{scalar}{v1,i1}
         \fmfv{d.shape=circle,d.size=5,label=$\Lambda_c^{*+}(P)$,label.angle=-40}{v1}
          \fmfv{label=$\theta_{12}$,label.angle=-160,label.dist=4mm}{v2}
        \fmf{plain}{v1,v2}
        \fmf{fermion,tension=50}{v2,o1}
        \fmf{scalar,tension=50}{v2,i2}
    \end{fmfgraph*}
    \vspace{0.2cm}
\end{fmffile}
\caption{\label{helicity1} The angle between the two pions in the rest frame of the intermediate resonance. In this rest frame, $\Lambda_c^+$ and $\pi^-$ are going back to back.}
\end{figure}

As anticipated one of important angles in the three-body decay of $\Lambda_c^* \to \Lambda_c \pi \pi$ is the angle between the two pions in the resonance's rest frame, $\theta_{12}$, which corresponds to the helicity angle.  
The definition of the angle between the two pions, $\theta_{12}$, in the rest frame of particle $\pi^-$(2) and $\Lambda_c^+$(3) is depicted in Fig.~\ref{helicity1}.
In the heavy quark limit, the heavy baryons are approximately treated as static particles, so that the angle between $\pi$ and $\Lambda_c$ is not relevant.
In addition to the standard Dalitz plot, we will also discuss angular correlation plots to describe three-body decays in the next section.

\begin{table}[b]
\caption{\label{2595} The contributions of the sequential processes and direct process to the total decay width of $\Lambda_c^{*}(2595)$ with the $\lambda$ and $\rho$ mode quark model assignments (in unit of MeV). }
\begin{ruledtabular}
\begin{tabular}{lcccc}
 \multirow{2}{*}{Contribution}  & {$\lambda$ mode} 	& \multicolumn{2}{c}{$\rho$ mode} &\multirow{2}{*}{Exp. [PDG]}\\	
 &$j=1$&$j=0$&$j=1$&\\
\hline
$\Sigma_c^{0} \pi^+	$	&  0.182				& -	&	0.770		& 0.624 (24\%)\\
$\Sigma_c^{++} \pi^-$	&  0.218 				& -	&	0.946		&0.624 (24\%)\\ 
Direct				&  0.004				& -	& 	0.004		&-\\ 
Interference			&  0.068				& -	&	-0.122		&- \\ \hline
$\Sigma_c^{+} \pi^0$	&  0.719 				& -	&	7.278		&- \\ 
Direct				&  0.005				& -	& 	0.005		&-\\ 
Interference			&  0.026				&-	&	-0.090		&- \\
\hline
$\Gamma_{\rm total}$	&  2.222	 			& -	&      8.791		&$2.6 \pm 0.6$\\
\end{tabular}
\end{ruledtabular}
\end{table}

\begin{figure}[t]
\centering
\includegraphics[scale=0.68]{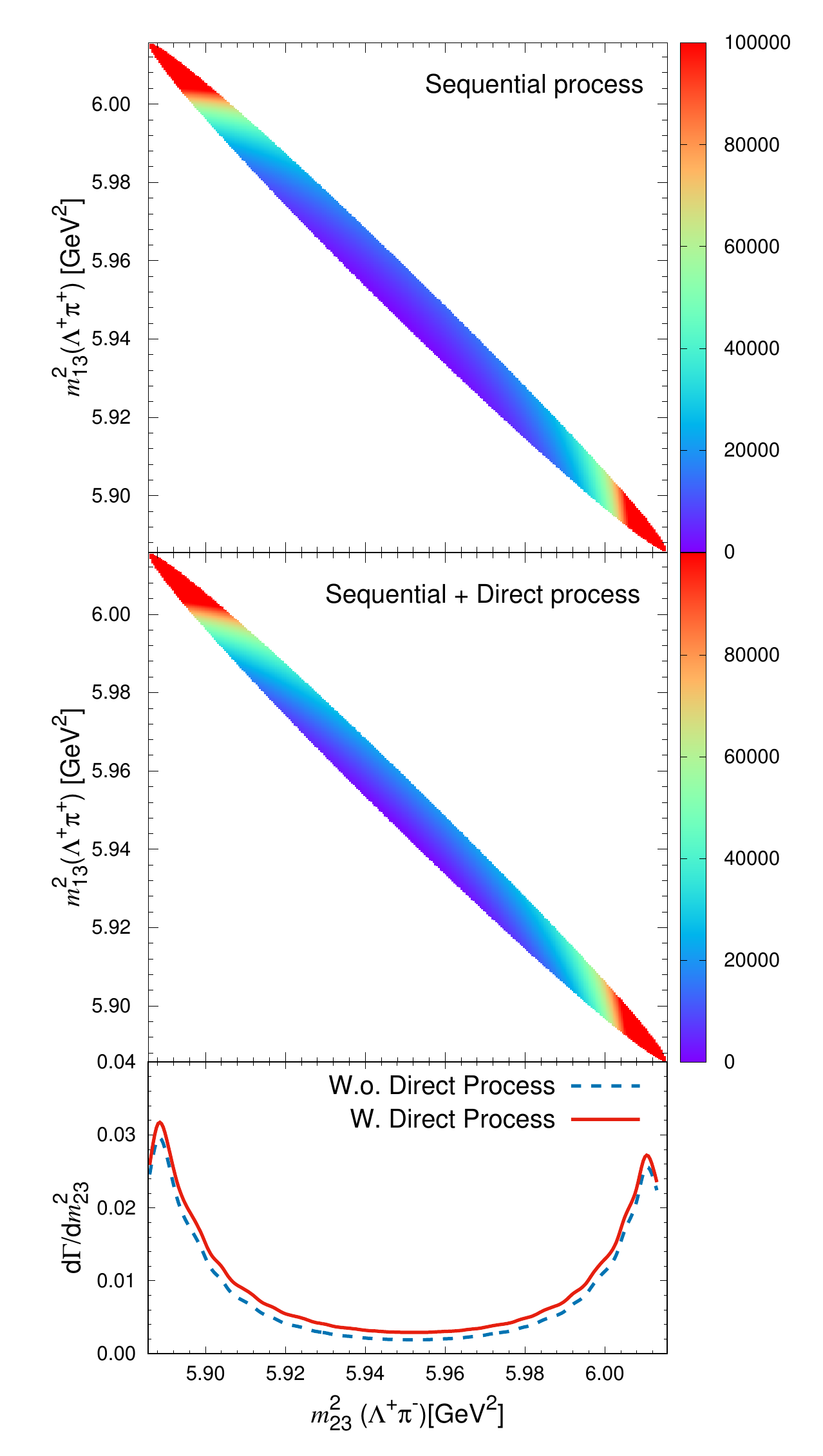}
\caption{\label{dalitz1} The Dalitz plots of  $\Lambda_c^{*}(2595)$ $ [\lambda\  {\rm mode}] \rightarrow \Lambda_c^+ \pi^+ \pi^-$ in $(m_{23}^2,m_{13}^2)$ plane and the invariant mass distribution of $m_{23}^2$ $(\Lambda_c^+\pi^-)$. The upper Dalitz plot is without the direct process and in the middle one the direct process has been included. }
\end{figure}

\section{Results and Discussions}
\subsection{$\Lambda_c^*(2595)$ Decay}

The first excited state $\Lambda_c^*(2595)$ has $J^P=1/2^-$ and $\Gamma=2.6 \pm 0.6$ MeV. 
This state is expected to be dominated by $p$-wave $\lambda$ mode excitation. 
The $\rho$ mode assignment overpredicts the data as shown in Table~\ref{2595}.
Although the absolute value itself is not enough to determine the structure, we will assume $\lambda$-mode dominance in this paper as we discussed previously~\cite{Arifi:2017sac}. 

Total decay width is well explained by the sequential process going through the $\Sigma_c(2455)$ open channel \cite{Nagahiro:2016nsx,Arifi:2017sac}. 
In the present work, we take into account the direct process and the interference among various terms. 
We neglect the contribution from the $\Sigma_c^*(2520)$ closed channel since the contribution is very small \cite{Arifi:2017sac}.
As shown in Table~\ref{2595}, the total decay width of $\Lambda_c^*(2595)$ is dominated by the neutral pion with the $\Sigma_c(2455)$ open channel and the contribution from the direct process is small.
This can be understood by considering the $s$-wave nature of $\Lambda_c^*\Sigma_c\pi$ coupling and $p$-wave of the direct coupling.

\begin{figure}[t]
\centering
\includegraphics[scale=0.68]{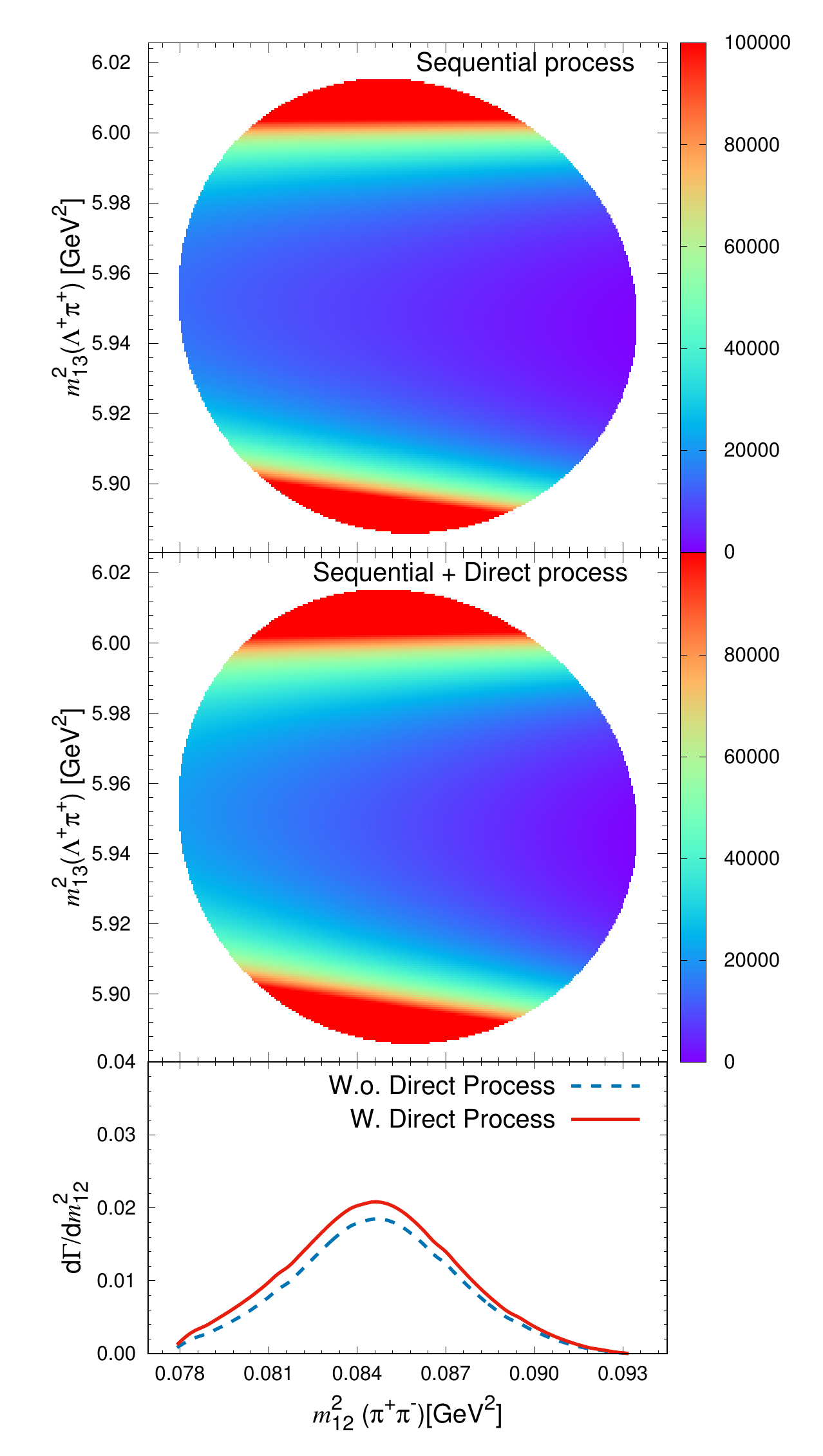}
\caption{\label{dalitz2} The Dalitz plots of $\Lambda_c^{*}(2595)$ $ [\lambda\  {\rm mode}] \rightarrow \Lambda_c^+ \pi^+ \pi^-$ in $(m_{12}^2,m_{13}^2)$ plane and the invariant mass distribution of $m_{12}^2$ $(\pi^+\pi^-)$. The upper Dalitz plot is without the direct process and in the middle one the direct process has been included. }
\end{figure}

Referring to Table~\ref{2595}, our result of the branching fraction of the $\Sigma_c(2455)^0\pi^+$ channel is given by
\begin{eqnarray}
\mathcal{B}(\Sigma_c(2455)^0\pi^+) &=& 0.082.
\end{eqnarray}
Our calculation is reasonably consistent with the recent branching fraction measurements from Belle, $\mathcal{B}(\Sigma_c(2455)^0\pi^+) = 0.125 \pm 0.035$ \cite{Niiyama:2017wpp}.
As shown in Table~\ref{2595}, the $\Sigma_c(2455)^+\pi^0$ neutral channel is dominant as compared to the other charged channel which leads to the isospin breaking effect as discussed in our previous work~\cite{Arifi:2017sac}.

In Fig.~\ref{dalitz1}, we show the Dalitz plots of $\Lambda_c^*(2595) \to \Lambda_c^+\pi^+\pi^-$ in $(m_{23}^2, m_{13}^2)$ plane, as well as the invariant mass plot with respect to $m_{23}^2$.
The upper figure shows the Dalitz plot without the direct process, while the middle one with direct process.  
As we discussed above, in this work we take into account the angular dependence properly, while we took angle average in our previous plot in Fig.~5 in Ref.~\cite{Arifi:2017sac}.
If we took the angle average, equal-strength lines parallel to either $m^2_{13}$ or $m^2_{23}$ axis.  
By including the angular dependence, they are inclined.

\begin{figure}[t]
\centering
\includegraphics[scale=0.55]{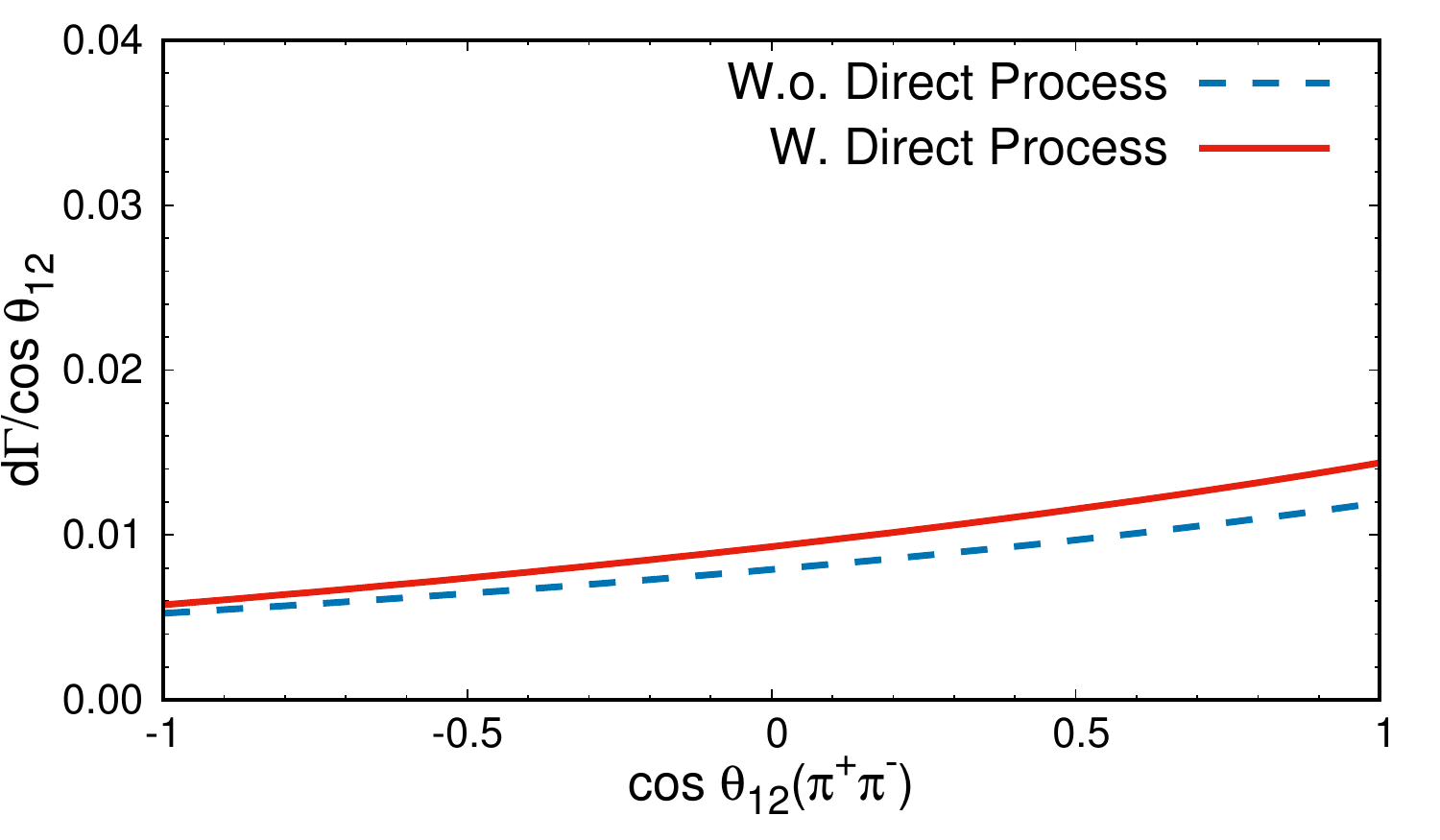}
\caption{\label{helicityangle1} The angular correlation between the two pions in the rest frame of particle 2 and 3 for the decay of $\Lambda_c^*(2595)$.}
\end{figure}

An alternative Dalitz plot in ($m^2_{12} , m^2_{13}$) plane shown in Fig.~\ref{dalitz2} is useful since it has a larger area that enables us to observe detailed structures such as the angular dependence clearly.
For the decay of $\Lambda_c^*(2595)$, the difference in the two plots is not very large, but we will see a clear difference for the case of $\Lambda_c^*(2625)$. 
We can make yet another plot; $\theta_{12}$ angular correlation, which is shown in Fig.~\ref{helicityangle1}.  
As in Figs.~\ref{dalitz1} and \ref{dalitz2}, the difference with and without direct process is not very large.   
What can be pointed here is that the angular correlation in Fig.~\ref{helicityangle1}, is 
due to the interference between $\Sigma_c(2455)$ in the different charged channels. 

\begin{table}[b]
\caption{ The contributions of the sequential processes and direct process to the total decay width of $\Lambda_c^{*}(2625)$ with the $\lambda$ and $\rho$ mode quark model assignments (in unit of MeV). }
\begin{ruledtabular}
\begin{tabular}{lcccc}
 \multirow{2}{*}{Contribution}  & {$\lambda$ mode}  &  \multicolumn{2}{c}{$\rho$ mode} & \multirow{2}{*}{Exp. [PDG]}  \\
	& $j=1$ & $j=1$ & $j=2$ & \\
\hline
$\Sigma_c^{0} \pi^+	$		& 0.037 	& 0.019 	& 0.034	&$< 0.050$ \\
$\Sigma_c^{++} \pi^-$		& 0.031 	& 0.016	& 0.028 	&$< 0.050$ \\
$\Sigma_c^{*0} \pi^+$ 		& 0.044 	& 0.197 	& 0.000 	&-\\
$\Sigma_c^{*++} \pi^-$ 		& 0.064	& 0.314 	& 0.000 	&-\\  
Direct					& 0.061	& 0.061	& 0.061	&- \\ 
Interference				& 0.090	& -0.166	& -0.011	&-  \\ \hline
$\Sigma_c^{+} \pi^0$		& 0.056 	& 0.029 	& 0.052 	&- \\
$\Sigma_c^{*+} \pi^0$ 		& 0.072 	& 0.325 	& 0.000 	&-\\
Direct					& 0.045	& 0.045	& 0.045	&- \\ 
Interference				& 0.070	& -0.130	&-0.009 	&- \\ 
\hline
$\Gamma_{\text{total}}$		& 0.570	& 0.710	& 0.200 	&$< 0.970$\\
\end{tabular}
\end{ruledtabular}
\label{result32}
\end{table}

\begin{figure}[t]
\centering
\includegraphics[scale=0.66]{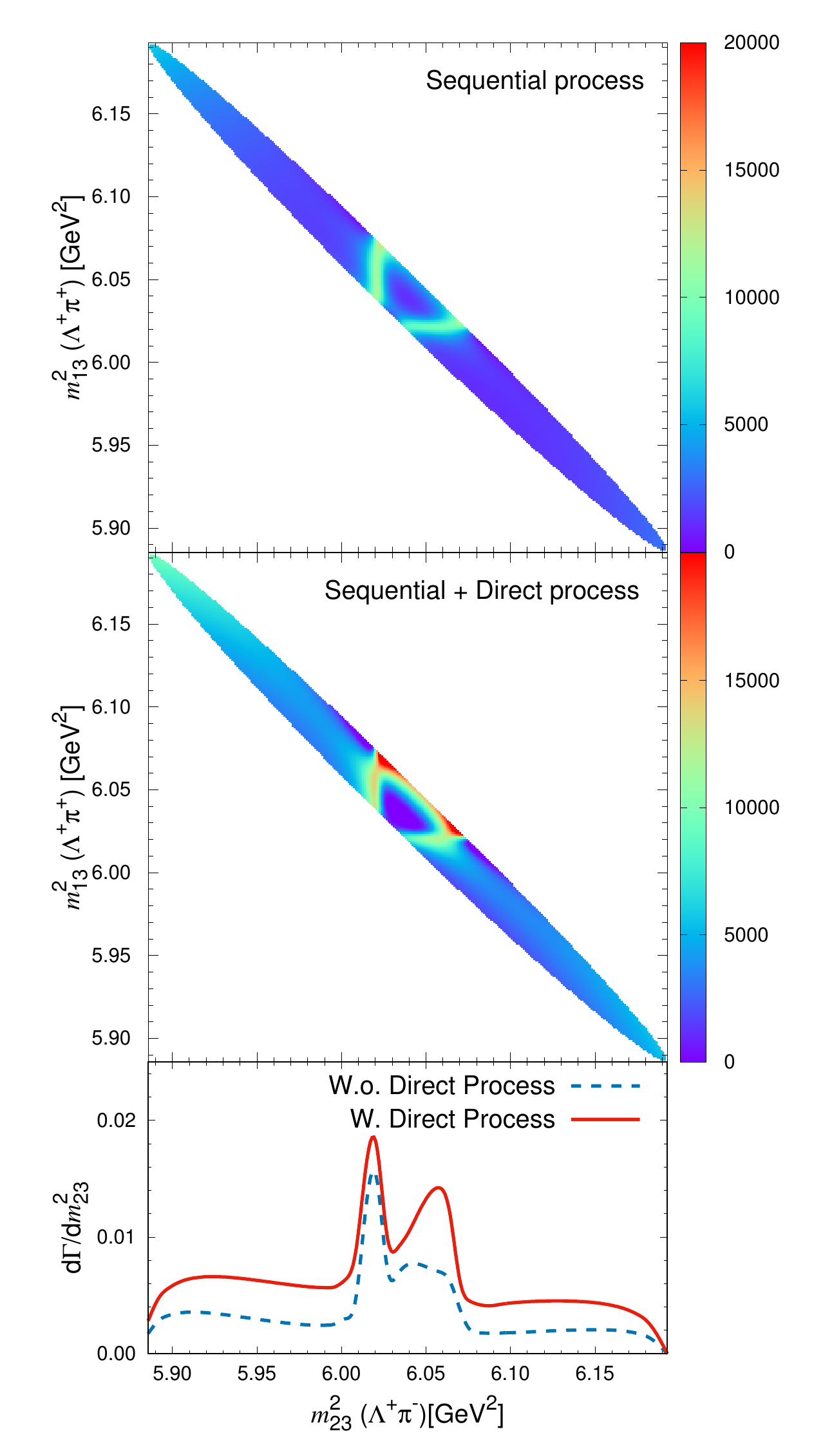}
\caption{\label{dalitz3} The Dalitz plots of $\Lambda_c^{*}(2625) $ $ [\lambda\  {\rm mode}] \rightarrow \Lambda_c^+ \pi^+ \pi^-$ in $(m_{23}^2,m_{13}^2)$ plane and the invariant mass distribution of $m_{23}^2$ $(\Lambda_c^+\pi^-)$. The upper Dalitz plot is without the direct process and in the middle one the direct process has been included. }
\end{figure}

\begin{figure}[t]
\centering
\includegraphics[scale=0.66]{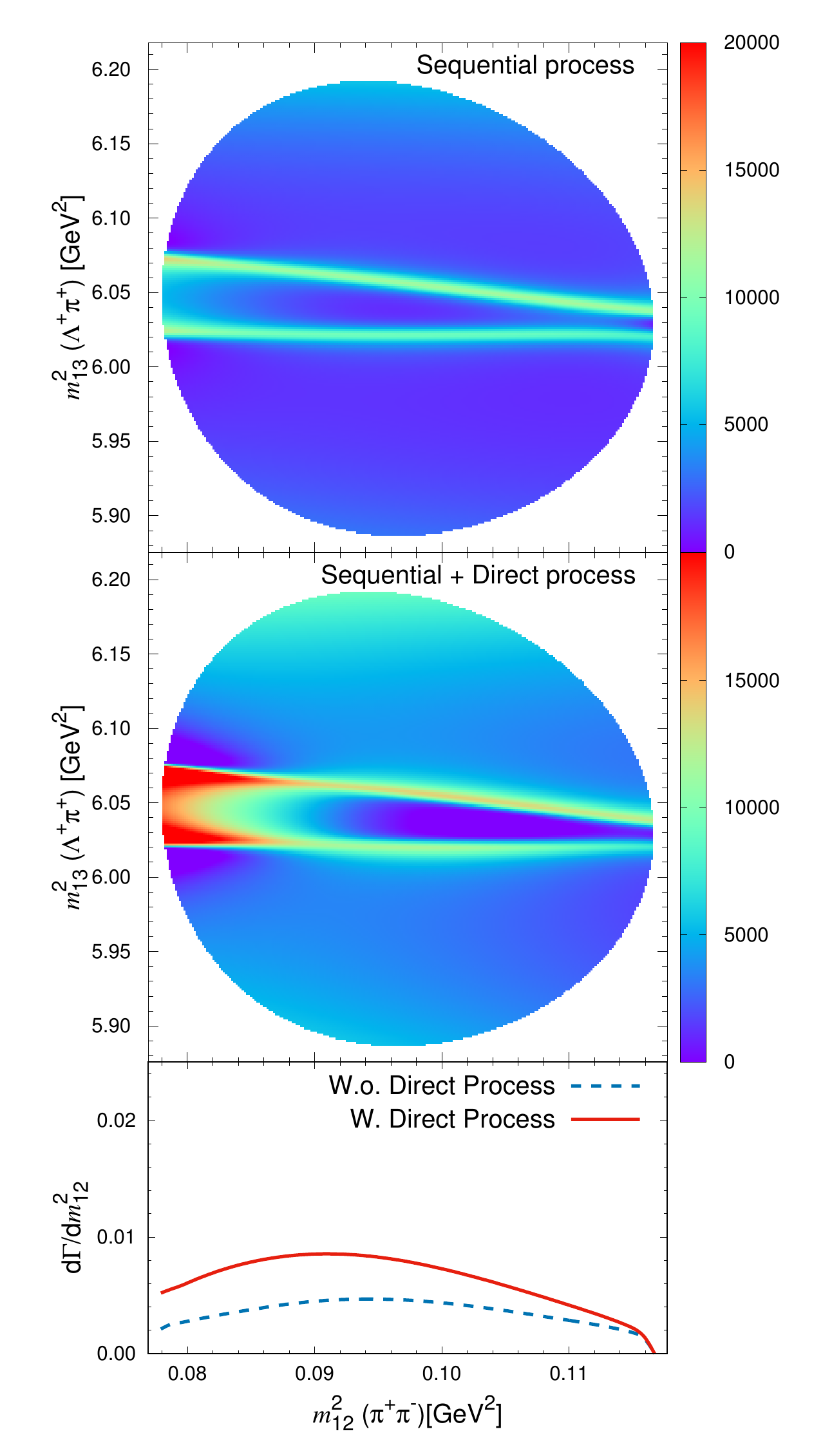}
\caption{\label{dalitz4}The Dalitz plots of $\Lambda_c^{*}(2625) $ $ [\lambda\  {\rm mode}] \rightarrow \Lambda_c^+ \pi^+ \pi^-$ in $(m_{12}^2,m_{13}^2)$ plane and the invariant mass distribution of $m_{12}^2$ $(\pi^+\pi^-)$. The upper Dalitz plot is without the direct process and in the lower one,  the direct process has been included.}
\end{figure}

\subsection{$\Lambda_c^*(2625)$ Decay}

The second resonance $\Lambda_c^*(2625)$ has $J^P=3/2^-$ and $\Gamma <0.97 $ MeV. 
Since this state is also expected to be dominated by $\lambda$ mode, we will mainly focus on $\lambda$ mode in this section.
For completeness, we provide the corresponding results for the other excitation modes: $\rho$ mode ($j=1$) and ($j=2$) in Appendix C and D.

According to PDG, the decay of this state is dominated by the non-resonant process.
In an attempt to explain the dominance of the non-resonant process,
we have considered the sequential process going through the $\Sigma_c^*(2520)$ closed channel \cite{Arifi:2017sac}.
In this work, we include the direct process in addition to the sequential ones.
Different from the case of $\Lambda_c^*(2595)$,  the direct process is important in this decay.
It has a considerable contribution as shown in Table~\ref{result32}. 
Because the $\Sigma_c(2455)$ open channel is suppressed due to the $d$-wave nature, the contribution of the sequential process becomes relatively smaller as compared to the direct process of $p$-wave nature.

By considering the sequential decay going through the $\Sigma_c^*(2520)$ closed channel and the direct process, 
we calculate the branching fraction of the $\Sigma_c(2455)^0\pi^+$ channel as 
\begin{eqnarray}
\mathcal{B}(\Sigma_c(2455)^0\pi^+) = 0.065.
\end{eqnarray}
The obtained value is consistent with the experimental data from Belle \cite{belle}.

For the decay of $\Lambda_c^*(2625)$, the Dalitz plot in $(m_{23}^2,m_{13}^2)$ plane is given in Fig.~\ref{dalitz3}.
One can notice that the Dalitz plot with and without the inclusion of the direct process are far different.
Non-trivial structures along the $\Sigma_c(2455)$ resonance's bands are due to the angular dependences from the interference between $\Sigma_c(2455)$ and the direct process. 
The effect of the angular dependences can be seen more clearly in $(m_{12}^2,m_{13}^2)$ plane as shown in Fig.~\ref{dalitz4}.
In this plot, the resonance's band is longer and then such angular dependences can be clearly seen so that it is easier to compare with the experimental data.
The plots show a characteristic pattern that the decay rate is accumulated on the left side of the resonance's band.  

\begin{figure}[t]
\centering
\includegraphics[scale=0.67]{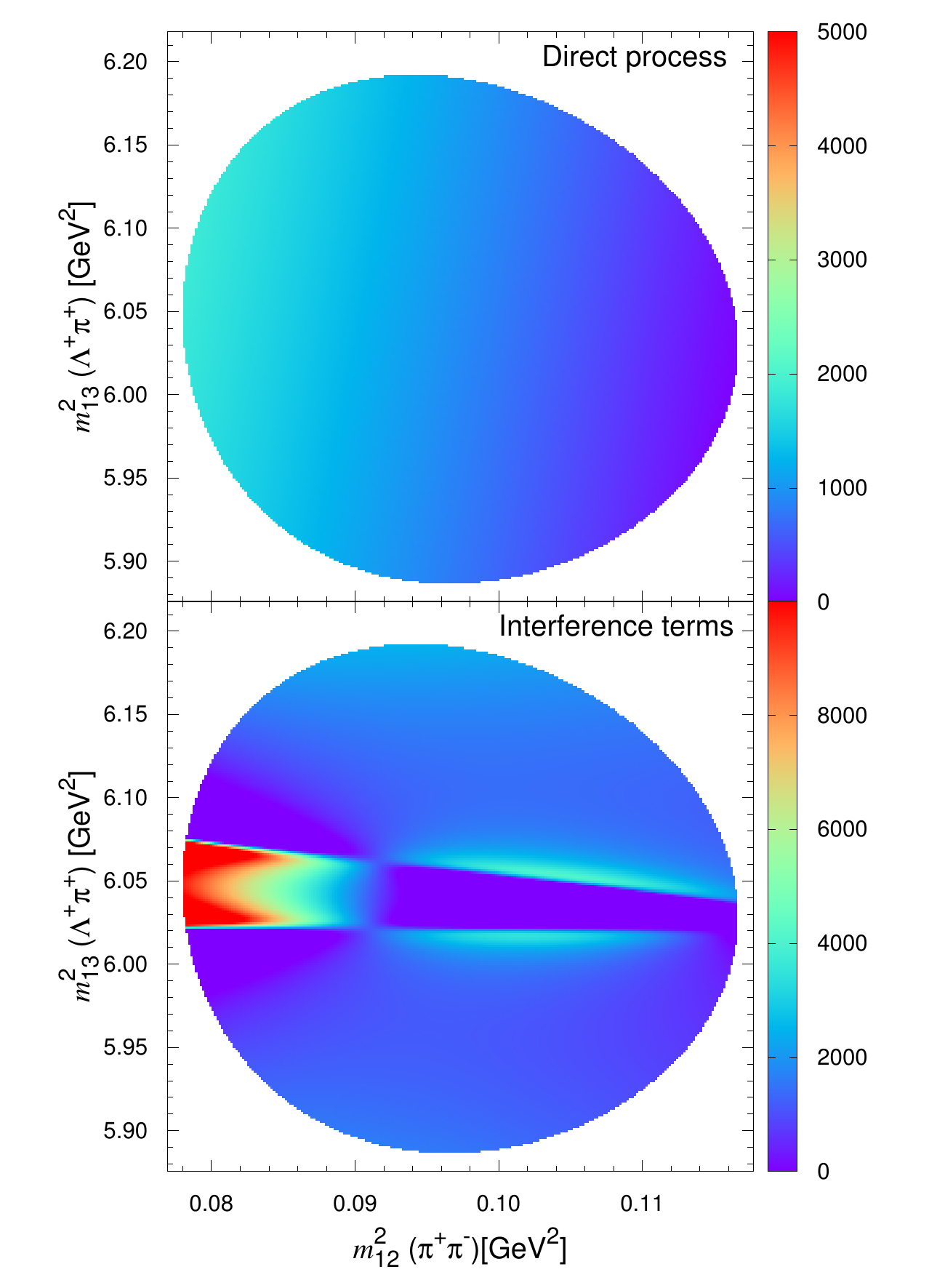}
\caption{\label{direct} The Dalitz plots of $\Lambda_c^{*}(2625) [\lambda\  {\rm mode}] \rightarrow \Lambda_c^+ \pi^+ \pi^-$ in $(m_{12}^2,m_{13}^2)$ plane. 
The direct process is exclusively shown  and the lower plot is the interference between the direct process and the sequential processes.}
\end{figure}

Now we would like to see various contributions separately. 
In Fig.~\ref{direct}, we show the contributions of the direct process and the interference between the direct and the other processes.
An asymmetric pattern is seen in the upper plot in Fig.~\ref{direct} which is solely from the direct process. 
In the lower plot, there is a large enhancement near the left boundary of the $\Sigma_c(2455)$ resonance's bands due to the interference between the direct process and the $\Sigma_c(2455)$ resonance.

Regarding the angular dependences, they are mainly originated from the direct process and the interference between the sequential and direct processes.
The interferences between $\Sigma_c(2455)^0$ and $\Sigma_c(2455)^{++}$ are suppressed because the overlapping region is small as compared to the total phase space.
It is depicted that there is no such a large interference along $\Sigma_c(2455)$ resonance's bands in the top plot of Fig.~\ref{dalitz4}.
Moreover, the interference between $\Sigma_c(2455)$ and $\Sigma_c^*(2520)$ are more suppressed since there is no overlapping region between the two resonance's peaks.
On the other hand, the interferences between the direct process and the other processes are dominant because the direct process contribution spreads over a wide region of the Dalitz plot.
In other words, the angular dependences are mostly originated from the direct process. 

\begin{figure}[t]
\centering
\includegraphics[scale=0.53]{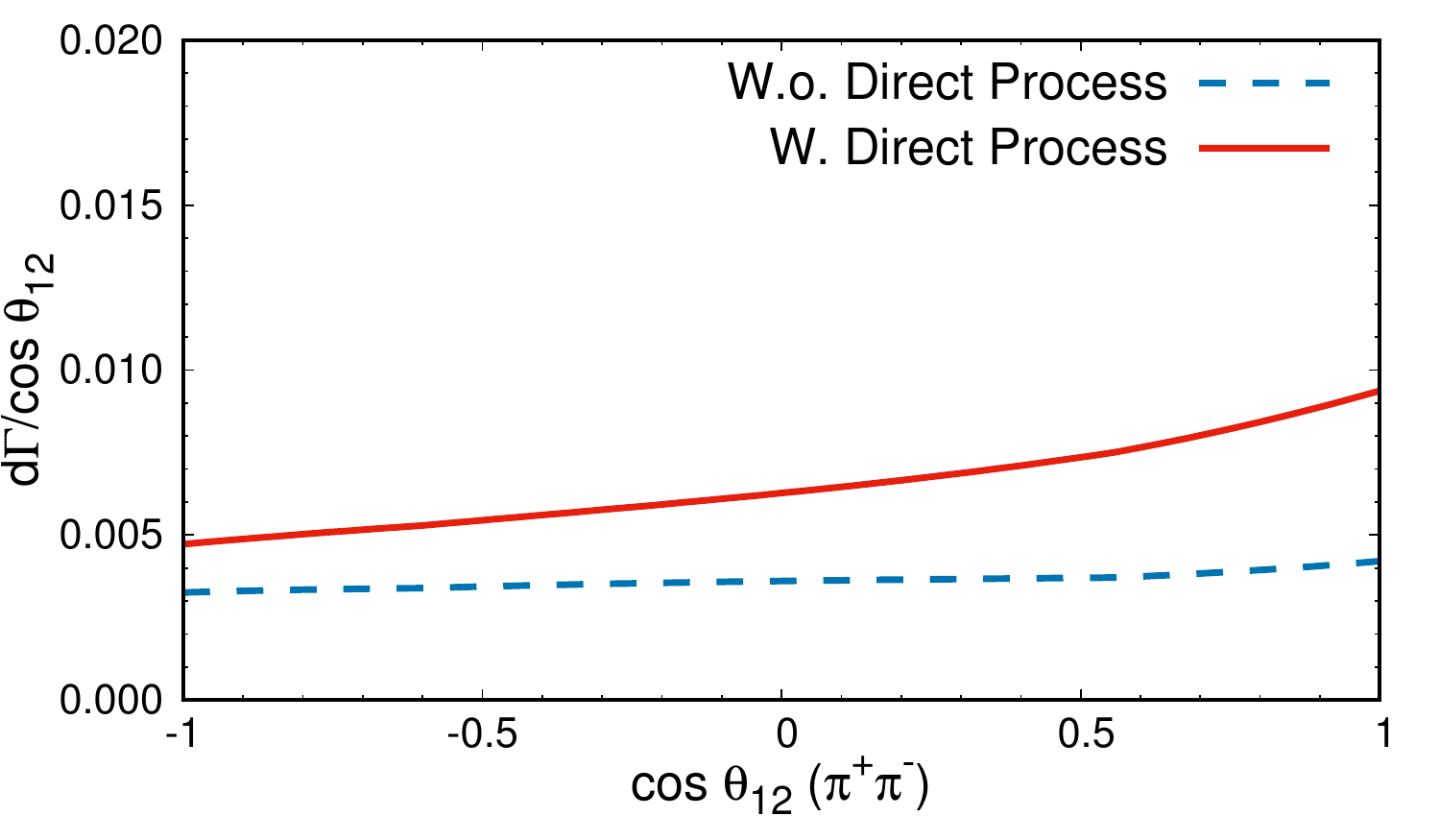}
\caption{\label{helicityangle3} The angular correlation between the two pions in the rest frame of particle 2 and 3 for the decay of $\Lambda_c^*(2625)$.}
\end{figure}

The main indication of the presence of the direct process is the asymmetric pattern in the angular correlations of the two pions as shown in Fig.~\ref{helicityangle3}. 
The angular correlation is approximately flat without the direct process which is contrast with the $\Lambda_c^*(2595)$ decay. 
For $\Lambda_c^*(2625)$, the angular dependence from the interferences between $\Sigma_c$'s and $\Sigma_c^*$'s is suppressed leading to the flat distribution to the angular correlation.
When the direct process is included, the asymmetric pattern appears. 
Therefore, this large asymmetric pattern can be a strong indication that can be measured in the experiment in the future.

Now it is interesting to see the effect of the direct process in $\rho$ mode excitations.
For $\rho$ mode ($j=1$), the $\Sigma_c^*(2520)$ contribution is large, while it is small for $\rho$ mode ($j=2$). 
The difference in their structure leads to various structures in Dalitz plot as we will discuss in detail in Appendix C and D, which is useful for the study of the internal structures.

\section{Summary}

We have investigated three-body decay of $\Lambda_c^*(2595)$ and $\Lambda_c^*(2625)$.
The direct two-pion emission process has been added in our model in addition to the sequential processes going through 
$\Sigma_c(2455)$ and $\Sigma_c^*(2520)$. 
We have estimated the strength of the direct coupling from the Yukawa coupling of the sequential decay by assuming the chiral partner structure for the relevant charmed baryon.
In this work, the angular dependences have been taken account properly in order to investigate the angular correlation between the two pions.
Finally, we have calculated the contribution of the each process to the total decay width.
In addition, we have provided Dalitz plots, invariant mass distributions, pion angular correlations in the presence of the direct process.
Those observables give the constrains to disentangle the internal structure of the charmed baryon. 

For $\Lambda_c^*(2595)$ decay, the direct process does not give a significant contribution.
On the other hand, for the case of $\Lambda_c^*(2625)$ with $\lambda$ mode assignment, 
the direct process contribution is large and it is found that the asymmetric pattern which occurs in the pion angular correlations.
This asymmetric pattern can be measured experimentally to indicate an important role of the direct process especially in the three-body decay of $\Lambda_c^*(2625)$.

\begin{acknowledgments}
This work is supported by a scholarship from the Ministry of Education, Culture, Science and Technology of Japan for A. J. A. 
and also Grants-in-Aid for Scientific Research (Grants No. JP17K05443 (C) for H. N.), (Grants No. JP17K05441 (C) for A. H.).
We would like to thank Kiyoshi Tanida for fruitful discussions.
\end{acknowledgments}

 % ---------------------------------------------------------------------------------------------------------
\appendix

\section{$\Lambda_c^*(2595)$ Decay Amplitude}

The amplitude of $\Lambda_c^*(2595)$ decay is expressed by
\begin{eqnarray}
\mathcal{T}_{1} (\Sigma_c^{0})     &=&  G_1\ \chi^\dagger_{\Lambda_c} (\boldsymbol{\sigma} \cdot {\bf p}_2)\ \chi_{\Lambda_c^*},\nonumber\\
\mathcal{T}_{2} (\Sigma_c^{++})   &=&  G_2\ \chi^\dagger_{\Lambda_c} (\boldsymbol{\sigma} \cdot {\bf p}_1)\ \chi_{\Lambda_c^*},\nonumber\\
\mathcal{T}_{3} ({\rm Direct})	    &=&  G_3\ \chi^\dagger_{\Lambda_c} \left( \boldsymbol{\sigma} \cdot ({\bf p}_1+{\bf p}_2)\right)\ \chi_{\Lambda_c^*},
\end{eqnarray}
where $G_i$ contains the information about the coupling constant, normalization, and the Breit-Wigner parameterization for each channel. 
For example, the first quantity is described as
\begin{eqnarray}
G_1 = \frac{g_a g_b \sqrt{2m_{\Lambda_c^*}}\sqrt{2m_{\Lambda_c}}}{(m_{23} - m_{\Sigma_c^0}) + i \Gamma_{\Sigma_c^0}/2}. \label{coupling1}
\end{eqnarray}
The other quantities can be derived accordingly.
For the direct process, the quantity is written as
\begin{eqnarray}
G_3 =\frac{g_b}{f_\pi} \sqrt{2m_{\Lambda_c^*}}\sqrt{2m_{\Lambda_c}}. \label{coupling2}
\end{eqnarray}
The estimation of the coupling strength is based on the discussion in section 2.

The squared amplitude with the spin sum of final states and spin average of initial spin states is given by
\begin{eqnarray}
\sum \overline{|\mathcal{T}|^2} &=& \sum |\mathcal{T}_1+\mathcal{T}_2+\mathcal{T}_3 |^2.
\end{eqnarray}
Therefore, the result can be written as
\begin{eqnarray}
|\mathcal{T}_1|^2 &=& |G_1|^2\ |{\bf p}_2|^2, \nonumber\\
|\mathcal{T}_2|^2 &=& |G_2|^2\ |{\bf p}_1|^2, \nonumber\\
|\mathcal{T}_3|^2 &=& |G_3|^2\ \left( |{\bf p}_1|^2 + |{\bf p}_2|^2 + 2 |{\bf p}_1| |{\bf p}_2| \cos\theta_{12} \right).\quad
\end{eqnarray}
From the equations above, we can observe that the direct process has $\cos\theta_{12}$ dependence which is the source of asymmetry in the angular correlation.
For the case of the sequential decay through $\Sigma_c(2455)$, no angular dependences appear resulting to a flat distribution.

For the case of angle average, the interference parts vanish. 
But, the interference parts turns out to be important if we take into account the angular dependence.
The interference between $\Sigma_c(2455)^{++}$ and $\Sigma_c(2455)^{0}$ is
\begin{eqnarray}
\mathcal{T}_1 \mathcal{T}_2^*	&=& G_1 G_2^*  \  |{\bf p}_1| |{\bf p}_2| \cos\theta_{12}. 
\end{eqnarray}
The interference between $\Sigma_c$'s is the main source of asymmetric pattern in $\Lambda_c^*(2595)$ decay.
The interference between direct process and sequential process also produce $\cos\theta_{12}$ dependence as given by
\begin{eqnarray}
\mathcal{T}_1 \mathcal{T}_3^*  &=& G_1 G_3^* \  \left(|{\bf p}_2|^2 + |{\bf p}_1| |{\bf p}_2| \cos\theta_{12} \right),\nonumber\\
\mathcal{T}_2 \mathcal{T}_3^*  &=& G_2 G_3^*\   \left(|{\bf p}_1|^2 + |{\bf p}_1| |{\bf p}_2| \cos\theta_{12}\right).
\end{eqnarray}
But, the contribution to the asymmetric pattern is small since the direct process is suppressed.

\section{$\Lambda_c^*(2625)$ Decay Amplitude}
The amplitude of $\Lambda_c^*(2625)$ decay is expressed by
\begin{eqnarray}
\mathcal{T}_{1} (\Sigma_c^{0})		&=& F_1\ \chi^\dagger_{\Lambda_c} (\boldsymbol{\sigma}  \cdot {\bf p}_2) \nonumber\\
							&& \times \left(\boldsymbol{\sigma} \cdot {\bf p}_1\ {\bf S} \cdot {\bf p}_1  -\frac{1}{3}   \boldsymbol{\sigma} \cdot {\bf S} |{\bf p}_1|^2\right) \chi_{\Lambda_c^*}, \quad\quad\nonumber\\
\mathcal{T}_{2} (\Sigma_c^{*0})    	&=&  F_2\ \chi^\dagger_{\Lambda_c} ({\bf S} \cdot {\bf p}_2)\ \chi_{\Lambda_c^*},\nonumber\\
\mathcal{T}_{3} (\Sigma_c^{++})   	&=&  F_3\ \chi^\dagger_{\Lambda_c} (\boldsymbol{\sigma}  \cdot {\bf p}_1) \nonumber\\
							&& \times \left(\boldsymbol{\sigma} \cdot {\bf p}_2\ {\bf S} \cdot {\bf p}_2  -\frac{1}{3}   \boldsymbol{\sigma} \cdot {\bf S} |{\bf p}_2|^2\right) \chi_{\Lambda_c^*},\nonumber\\
\mathcal{T}_{4} (\Sigma_c^{*++})  	&=&  F_4\ \chi^\dagger_{\Lambda_c} ({\bf S} \cdot {\bf p}_1) \chi_{\Lambda_c^*}, \nonumber\\
\mathcal{T}_{5} ({\rm Direct})	    	&=&  F_5\ \chi^\dagger_{\Lambda_c} \left( {\bf S} \cdot ({\bf p}_1+{\bf p}_2)\right) \chi_{\Lambda_c^*}.\hspace{2.4cm}
\end{eqnarray}
where $F_i$ has similar structures as in Eqs. (\ref{coupling1}) and (\ref{coupling2}).

\begin{figure}[b]
\centering
\includegraphics[scale=0.68]{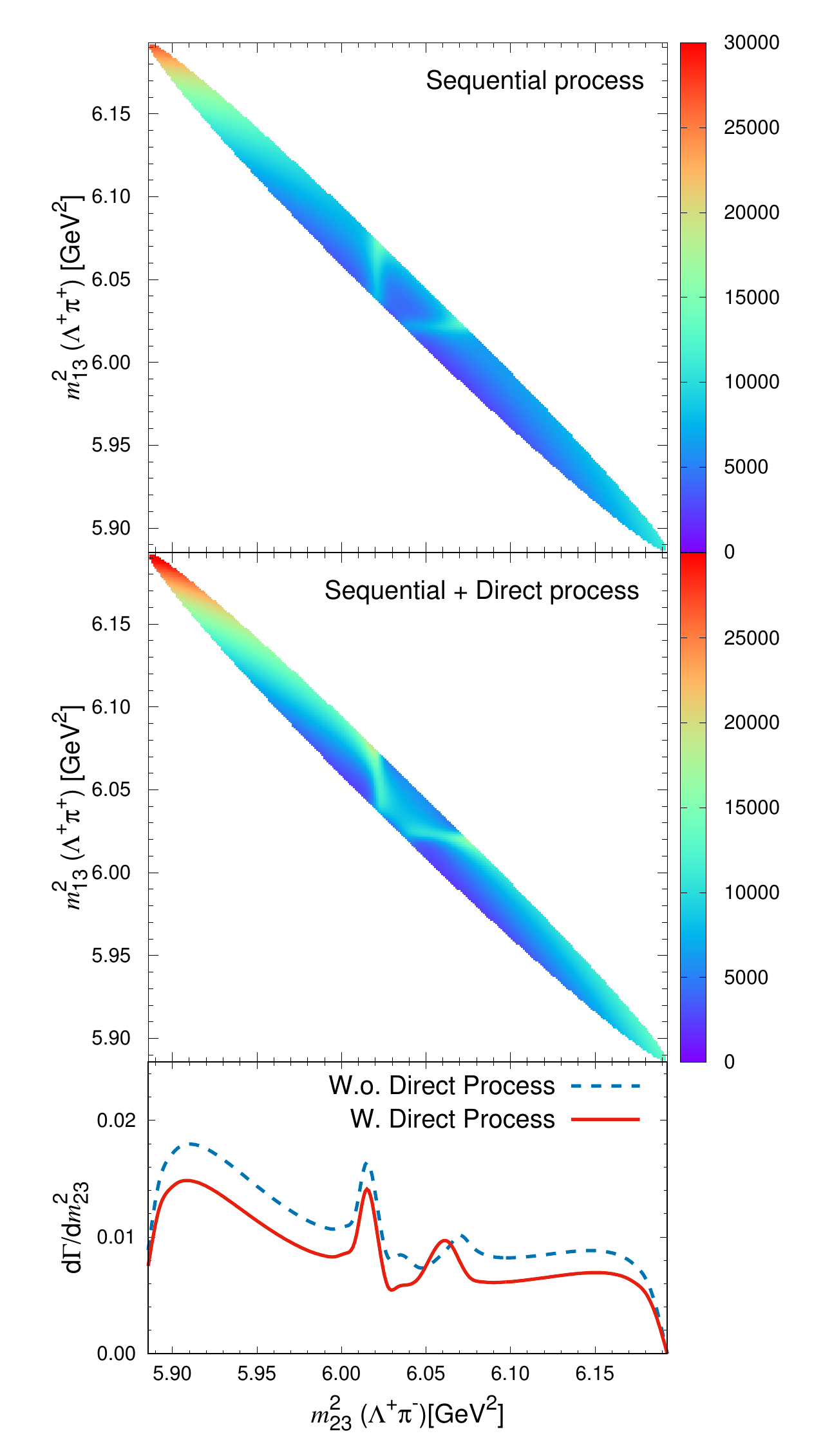}
\caption{\label{dalitz_rho1} The Dalitz plots of $\Lambda_c^{*}(2625) $ $ [\rho\  {\rm mode}, j=1] \rightarrow \Lambda_c^+ \pi^+ \pi^-$ in $(m_{23}^2,m_{13}^2)$ plane and the invariant mass distribution of $m_{23}^2$ $(\Lambda_c^+\pi^-)$. The upper Dalitz plot is without the direct process and in the middle one the direct process has been included. }
\end{figure}

The squared amplitude with the spin sum of final state and spin average of initial spin state is given by
\begin{eqnarray}
\sum \overline{|\mathcal{T}|^2} &=& \sum |\mathcal{T}_1+\mathcal{T}_2+\mathcal{T}_3+\mathcal{T}_4+\mathcal{T}_5 |^2.
\end{eqnarray}
Then, the squared amplitude reads
\begin{eqnarray}
|\mathcal{T}_1|^2 &=& \frac{1}{3}\ |F_1|^2\ |{\bf p}_1|^4\ |{\bf p}_2|^2, \nonumber\\
|\mathcal{T}_2|^2 &=& \frac{1}{3}\ |F_2|^2\ |{\bf p}_2 |^2, \nonumber\\
|\mathcal{T}_3|^2 &=& \frac{1}{3}\ |F_3|^2\ |{\bf p}_2|^4\ |{\bf p}_1|^2, \nonumber\\
|\mathcal{T}_4|^2 &=& \frac{1}{3}\ |F_4|^2\  |{\bf p}_1 |^2,\  \nonumber\\
|\mathcal{T}_5|^2 &=& \frac{1}{3}\ |F_5|^2\ \left( |{\bf p}_1|^2 + |{\bf p}_2|^2 + 2 |{\bf p}_1| |{\bf p}_2| \cos\theta_{12} \right).\quad
\end{eqnarray}
If the initial particle is polarized, the sequential decay process going through $\Sigma_c^*(2520)$ produces the angular dependences such as
\begin{eqnarray}
|\mathcal{T}_2|^2 (h=+1/2) &=& \frac{1}{3}\ |F_2|^2\ |{\bf p}_2 |^2 \left(3\cos^2\theta_{12} +1 \right),\quad\\
|\mathcal{T}_2|^2 (h=+3/2) &=& |F_2|^2\  |{\bf p}_2 |^2 \sin^2\theta_{12}. 
\end{eqnarray}
But, they will vanish in unpolarized case. 
The interferences between $\Sigma_c(2455)^{(0,++)}$ and $\Sigma_c^*(2520)^{(0,++)}$ are given by
\begin{eqnarray}
\mathcal{T}_1 \mathcal{T}_2^*  &=& \frac{1}{6}\ F_1 F_2^*\ |{\bf p}_1 |^2\ |{\bf p}_2|^2 \Bigl(3 \cos^2\theta_{12} -1 \Bigr), \nonumber\\
\mathcal{T}_3 \mathcal{T}_4^*  &=& \frac{1}{6}\ F_3 F_4^* \   |{\bf p}_1 |^2\ |{\bf p}_2|^2 \Bigl(3 \cos^2\theta_{12} -1 \Bigr),  \nonumber\\
\mathcal{T}_1 \mathcal{T}_4^*  &=& \frac{1}{3}\ F_1 F_4^* \  |{\bf p}_1|^3 |{\bf p}_2| \cos\theta_{12},\nonumber\\
\mathcal{T}_2 \mathcal{T}_3^*  &=& \frac{1}{3}\ F_2 F_3^* \  |{\bf p}_2|^3 |{\bf p}_1| \cos\theta_{12}.
\end{eqnarray}
The interference between $\Sigma_c(2455)$ and $\Sigma_c^*(2520)$ in the same charged channel gives the symmetric pattern,
but the interference between them in the different charged channel gives the asymmetric pattern.
The interference between $\Sigma_c(2455)^{++}$ and $\Sigma_c(2455)^{0}$ is written as
\begin{eqnarray}
\mathcal{T}_1 \mathcal{T}_3^*	&=& \frac{1}{3}\ F_1 F_3^*  \ |{\bf p}_1|^3 |{\bf p}_2|^3 \left( 3 \cos^2\theta_{12} - 2 \right) \cos\theta_{12}. \quad\quad
\end{eqnarray}
The interference between $\Sigma_c(2520)^{*++}$ and $\Sigma_c(2520)^{*0}$ is
\begin{eqnarray}
\mathcal{T}_2 \mathcal{T}_4^*  &=& \frac{1}{6}\ F_2 F_4^* \  |{\bf p}_1| |{\bf p}_2| \cos\theta_{12}.
\end{eqnarray}
The interferences between the direct process and the sequential processes are given by
\begin{eqnarray}
\mathcal{T}_1 \mathcal{T}_5^*  &=& \frac{1}{2}\ F_1 F_5^* \   |{\bf p}_1|^3|{\bf p}_2| \nonumber\\
		& & \times \Bigl( 2 \cos\theta_{12} + \frac{|{\bf p}_2|}{|{\bf p}_1|}\Bigl(3 \cos^2\theta_{12} -1 \Bigr) \Bigr),\nonumber\\
\mathcal{T}_3 \mathcal{T}_5^*  &=& \frac{1}{2}\ F_3 F_5^* \   |{\bf p}_2|^3 |{\bf p}_1| \nonumber\\
		& & \times \Bigl( 2 \cos\theta_{12} + \frac{|{\bf p}_1|}{|{\bf p}_2|}\Bigl(3 \cos^2\theta_{12} -1 \Bigr) \Bigr),\nonumber\\
\mathcal{T}_2 \mathcal{T}_5^*  &=& \frac{1}{6}\ F_2 F_5^* \  \left(|{\bf p}_2|^2 + |{\bf p}_1| |{\bf p}_2| \cos\theta_{12} \right), \nonumber\\
\mathcal{T}_4 \mathcal{T}_5^* &=& \frac{1}{6}\ F_4 F_5^*\   \left(|{\bf p}_1|^2 + |{\bf p}_1| |{\bf p}_2| \cos\theta_{12} \right).
\end{eqnarray}

\begin{figure}[b]
\centering
\includegraphics[scale=0.53]{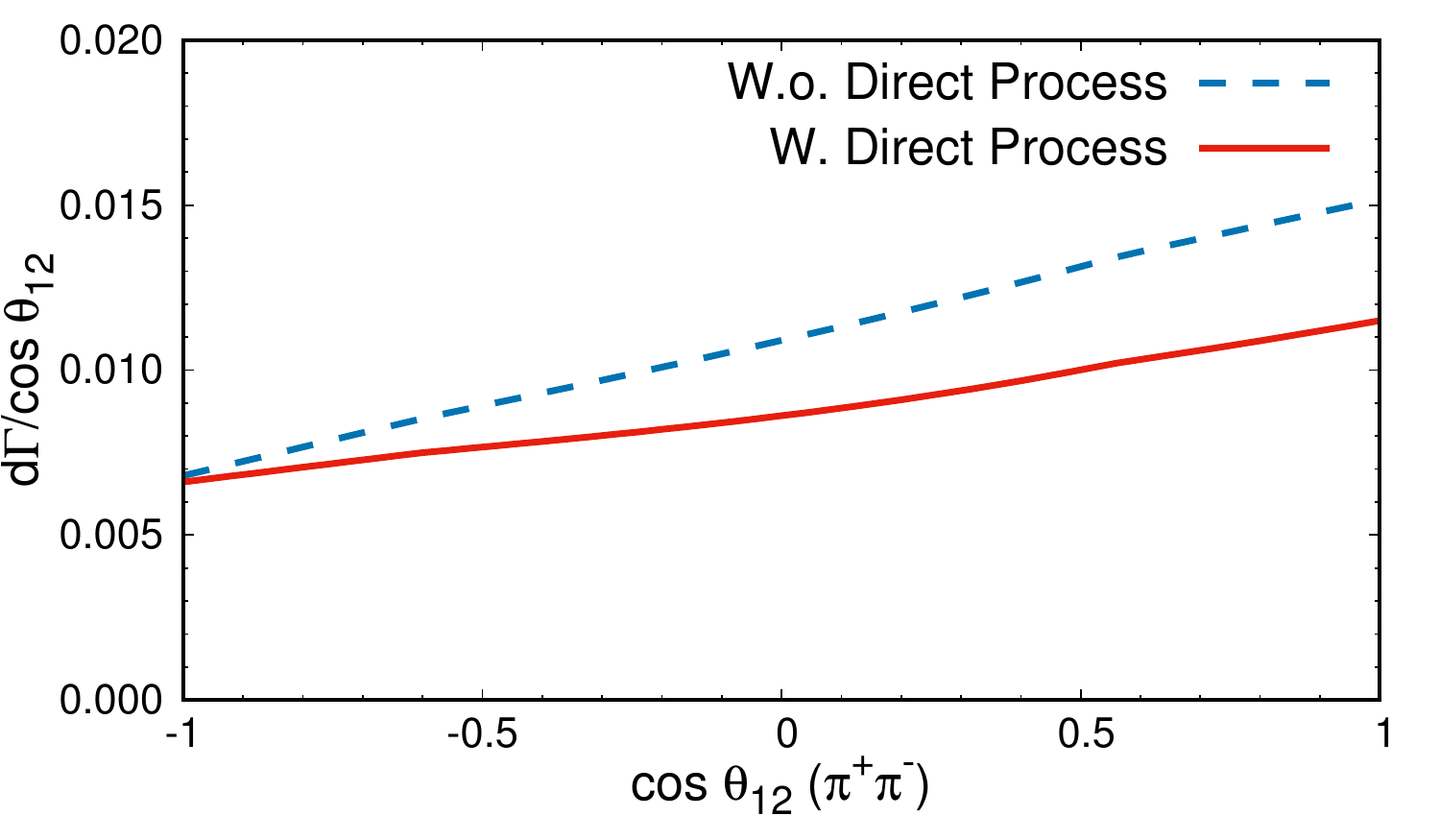}
\caption{\label{helicity_rho1} The angular correlation between the two pions in the rest frame of particle 2 and 3 for the decay of $\Lambda_c^*(2625)$ with $\rho$ mode ($j$=1).}
\end{figure}

\begin{figure}[b]
\centering
\includegraphics[scale=0.68]{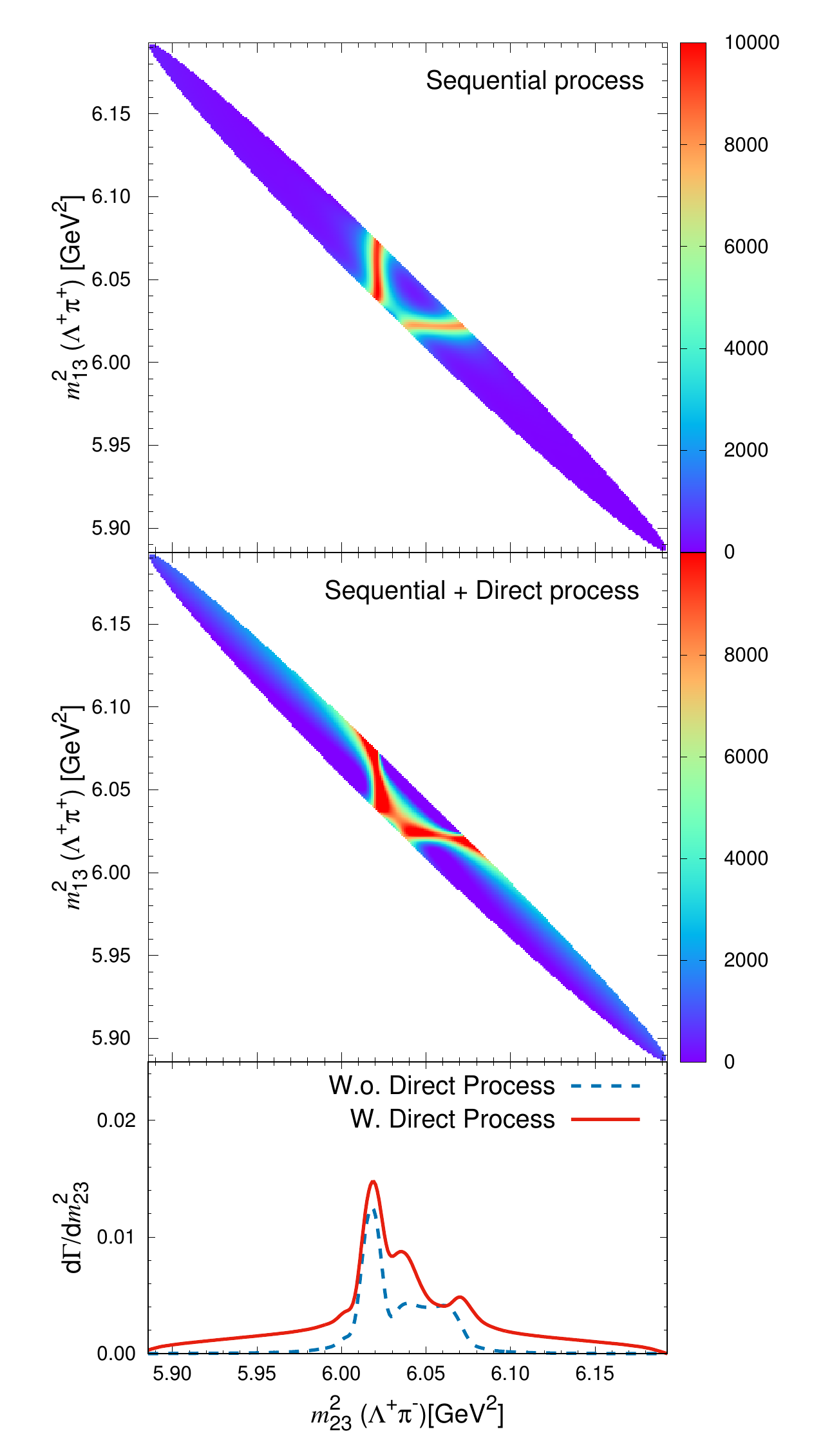}
\caption{\label{dalitz_rho2} The Dalitz plots of $\Lambda_c^{*}(2625) $ $ [\rho\  {\rm mode}, j=2] \rightarrow \Lambda_c^+ \pi^+ \pi^-$ in $(m_{23}^2,m_{13}^2)$ plane and the invariant mass distribution of $m_{23}^2$ $(\Lambda_c^+\pi^-)$. The upper Dalitz plot is without the direct process and in the middle one the direct process has been included.}
\end{figure}

\begin{figure}[b]
\centering
\includegraphics[scale=0.53]{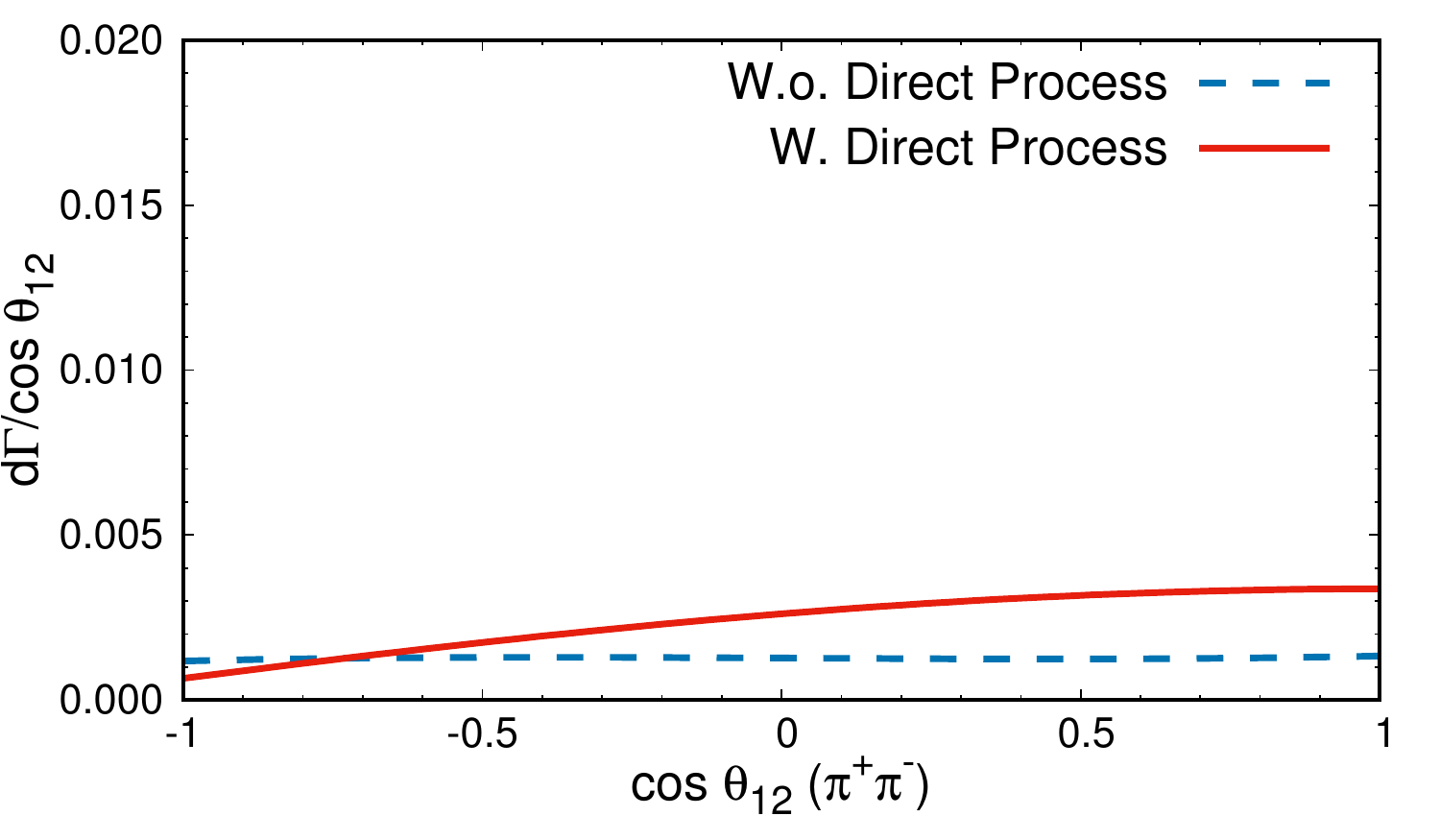}
\caption{\label{helicity_rho2} The angular correlation between the two pions in the rest frame of particle 2 and 3 for the decay of $\Lambda_c^*(2625)$ with $\rho$ mode ($j$=2).}
\end{figure}

\section{$\Lambda_c^*(2625)$ [$\rho$ mode ($j$=1)]}

For the decay of $\Lambda_c^*(2625)$ with $\rho$ mode $(j =1)$, the $\Sigma_c^*(2520)$ closed channel is very large as shown in Table~\ref{result32}. 
One can observe that the large contribution of $\Sigma_c^*(2520)$ is reflected as the large enhancement near the boundaries of the Dalitz plots and the invariant mass distribution in Fig.~\ref{dalitz_rho1}. 
For this assignment, the contribution of the direct process is relatively smaller than the $\Sigma_c^*(2520)$ closed channel. 
In other words, the non-resonant contribution for this decay is mostly originated from the $\Sigma_c^*(2520)$ closed channel.

When we take into account the direct process, the interferences between the direct process and the sequential processes give destructive patterns as tabulated in Table~\ref{result32}.
The suppression in the invariant mass distribution and the angular correlations are seen in Figs.~\ref{dalitz_rho1} and \ref{helicity_rho1} respectively.
Due to the destructive interference, the structure along the $\Sigma_c(2455)$ resonance's band has different shape in comparison with $\lambda$ mode case in Fig.~\ref{dalitz3}. 
There is no enhanced part in the end of the resonance's band in $\rho$ mode case.

The interferences between $\Sigma_c^*(2520)$'s become much enhanced because of the strong coupling to $\Sigma_c^*(2520)$.
As a consequence, large asymmetric pattern can be seen in angular correlations without direct process as shown in Fig.~\ref{helicity_rho1}.
This is because the interferences between $\Sigma_c^*(2520)$'s giving $\cos\theta_{12}$ dependence is now significant.

For $\Lambda_c^*(2625)$ with $\rho$ mode $(j=1)$, the indication of the direct process is not clear as $\lambda$ mode case, since $\cos\theta_{12}$ dependence is now dominated from the interference between $\Sigma_c^*(2520)^{0}$ and $\Sigma_c^*(2520)^{++}$, not from the direct process. 
Even without the direct process, the angular correlation has already exhibited the asymmetric pattern. 
In this assignment, the angular correlation with the direct process shows similar structure to $\lambda$ mode case.
The difference is the source of generating the asymmetric pattern in angular correlation.
One possible way to distinguish either $\lambda$ or $\rho$ mode $(j=1)$ is by measuring the invariant mass distribution,
they really exhibit different structures.

\section{$\Lambda_c^*(2625)$ [$\rho$ mode ($j$=2)]}

In the last assignment for $\Lambda_c^*(2625)$, namely $\rho$ mode $(j=2)$, 
we can see that the contribution from the $\Sigma_c^*(2520)$ closed channel is extremely small as shown in Table~\ref{result32}. 
It is opposite to the fact that the non-resonant contribution is considerably large according to PDG.
In this point of view, we can rule out the possibility of $\Lambda_c^*(2625)$ to be $\rho$ mode with $j=2$. 
Nevertheless, the effect of the inclusion of the direct process is still of the interest to discuss.
In this case, any asymmetric pattern can be roughly considered as the consequence of the direct process completely due to small contribution from $\Sigma_c^*(2520)$.

The Dalitz plots in $(m_{23}^2,m_{12}^2)$ plane and the invariant mass distribution of $\Lambda_c^+\pi^-$ with and without the direct process are given in Fig.~\ref{dalitz_rho2}.
The destructive interferences between the direct process and the sequential processes produce the similar angle dependences with $\rho$ mode $(j=1)$ in the $\Sigma_c(2455)$ resonance's band.
Due to angle dependences coming from the destructive interferences, a fake peak appears between the $\Sigma_c(2455)^0$ and the $\Sigma_c(2455)^{++}$ resonance's peaks in invariant mass distribution as shown in Fig. ~\ref{dalitz_rho2}.
The angular correlation also shows the asymmetric pattern due to the presence of the direct process as shown in Fig.~\ref{helicity_rho2}.

\nocite{*}
\bibliography{apssamp}% Produces the bibliography via BibTeX.

\end{document}